\documentclass[twocolumn,showpacs,prl]{revtex4}
\usepackage{graphicx}
\usepackage{dcolumn}
\usepackage{bm}

\begin{document}

\title{Trapped Fermions across a Feshbach resonance with population imbalance}
\author{W. Yi and L.-M. Duan}
\address{FOCUS center and MCTP, Department of Physics, University of Michigan, Ann
Arbor, MI 48109}

\begin{abstract}
We investigate the phase separation of resonantly interacting
fermions in a trap with imbalanced spin populations, both at zero
and at finite temperatures. We directly minimize the
thermodynamical potential under the local density approximation
instead of using the gap equation, as the latter may give unstable
solutions. On the BEC side of the resonance, one may cross three
different phases from the trap center to the edge; while on the
BCS side or at resonance, typically only two phases show up. We
compare our results with the recent experiment, and the agreement
is remarkable. 
\end{abstract}

\maketitle

The recent experimental realizations of condensation and
superfluidity in Fermi gases near a Feshbach resonance have
provided a new platform for controlled study of many-body physics
in the strongly correlated region \cite {1}. A significant advance
in this direction is illustrated by two very recent experiments
\cite{2,3}, which study the fermi superfluidity with controlled
population imbalance of different spin components. With increasing
population imbalance, a phase transition from the superfluid to
the normal state has been observed \cite{2}, and phase separation
of the Fermi gas in the trap has been identified \cite{2,3}.

The Cooper pairing in the Fermi gas, which is essential for the
Fermi condensate and the superfluidity, requires an equal number
of atoms from both spin components. If the atomic gas has
imbalanced population in its two spin states, inevitably some of
the atoms will be unpaired, which triggers the competition between
the Cooper pairing and the population imbalance. Such a
competition may lead to various new phenomena \cite {4,5,6,7,8},
which have raised strong interest recently \cite{6,7,8}.

In this paper, we present our study on the trapped fermions across
a Feshbach resonance with population imbalance. Compared with
other recent theoretical works \cite{6,7,8}, we demonstrate the
following results: (i) We consider a wide resonance with
parameters corresponding to the recent $^{6}Li$ experiment, and
calculate the pairing gap and the density distribution, which can
be directly compared with the experiment. (ii) We take into
account the trapping potential, and show that the trap favors the
phase separation, which is critical for the understanding of the
recent experiments \cite{2,3}. Some of the co-existing or unstable
phases in a homogeneous gas \cite{8} necessarily become phase
separated even if a very weak trap is turned on. (iii) We
investigate the effect of the population imbalance both at zero
and at finite temperatures. At zero temperature, the BCS
superfluid state requires
an equal number of atoms from both spin components; but at finite temperature $%
T $ it allows population imbalance carried by the quasiparticle
excitations. This key difference significantly relaxes the
competition between the population imbalance and the Cooper
pairing at finite $T$, which helps to explain the robustness of
the superfluid state in the recent experiment \cite {2}. (iv) We
directly minimize the thermodynamical potential to find out the
stable configuration of the system, and caution the use of the gap
equation in the case of population imbalance. In the region with
competition between different phases, the gap equation often gives
incorrect results or unstable solutions.

To describe the fermions across a Feshbach resonance, we take the following
two-channel Hamiltonian \cite{9,10,11}:
\begin{eqnarray}
H &=&\sum_{\mathbf{k},\sigma }(\epsilon _{\mathbf{k}}-\mu _{\sigma })a_{%
\mathbf{k},\sigma }^{\dag }a_{\mathbf{k},\sigma } \nonumber\\
&+&\sum_{\mathbf{q}}(\epsilon _{\mathbf{q}}/2+\gamma -\mu
_{\uparrow }-\mu
_{\downarrow })b_{\mathbf{q}}^{\dag }b_{\mathbf{q}} \nonumber\\
&+&\alpha/\sqrt{\mathcal{V}} \sum_{\mathbf{q},\mathbf{k}}(b_{\mathbf{q}}^{\dag }a_{\mathbf{q}/2-%
\mathbf{k},\downarrow }a_{\mathbf{q}/2+\mathbf{k},\uparrow }+h.c.), \\
&+&U/\mathcal{V}\sum_{\mathbf{q},\mathbf{k},\mathbf{k^{\prime }}}a_{\mathbf{q}/2+\mathbf{%
k},\uparrow }^{\dag }a_{\mathbf{q}/2-\mathbf{k},\downarrow }^{\dag }a_{%
\mathbf{q}/2-\mathbf{k^{\prime }},\downarrow }a_{\mathbf{q}/2+\mathbf{%
k^{\prime }},\uparrow }\nonumber
\end{eqnarray}
where $\epsilon _{\mathbf{k}}=k^{2}/(2m)$ ($m$ is the atom mass
and $\hbar =1 $), $\mu _{\sigma }$ is the chemical potential for
the spin-$\sigma $ component ($\sigma =\uparrow ,\downarrow $
labels the two spin states), $\mathcal{V}$ is the quantization
volume, $a_{ \mathbf{k},\sigma }^{\dag }$ and
$b_{\mathbf{q}}^{\dag }$ are the creation operators for the
fermionic atoms (open channel) and the bosonic molecules (closed
channel), respectively. The bare atom-molecule coupling rate
$\alpha $, the bare background scattering rate $U$, and the bare
detuning $\gamma $ are connected with the physical ones $\alpha
_{p},U_{p},\gamma _{p}$ through the standard renormalization
relations \cite{11}, and the values of $\alpha _{p},U_{p},\gamma
_{p}$ are determined from the scattering parameters (see, e.g.,
the explicit expressions in Ref. \cite{12}.). We take the local
density approximation so that $\mu _{\uparrow }=\mu
_{\mathbf{r}}+h,$ $\mu
_{\downarrow }=\mu _{\mathbf{r}}-h$, $\mu _{\mathbf{r}}=\mu -V\left( \mathbf{%
r}\right) $, where $V\left( \mathbf{r}\right) $ is the external trap
potential (slowly varying in $\mathbf{r}$), and $\mu ,h$ are determined from
the total atom number $N=N_{\uparrow }+N_{\downarrow }$ and the population
imbalance $\beta =\left| N_{\uparrow }-N_{\downarrow }\right| /N$ through
the number equations below.

For simplicity, we first take a mean-field approach by assuming $\langle b_{%
\mathbf{q}}\rangle =\langle b_{0}\rangle \delta
_{\mathbf{q0}}=-\alpha/\sqrt{\mathcal{V}}
\sum_{\mathbf{k}}\left\langle a_{-\mathbf{k},\downarrow }a_{\mathbf{k%
},\uparrow }\right\rangle /\left( \gamma -2\mu _{\mathbf{r}}\right) $ (the
last equality comes from the Heisenberg equation for the operator $b_{%
\mathbf{q}}$). At this level, we neglect the pair/molecule
fluctuation, an approximation that is valid near zero temperature
or in the BCS limit. Later we will discuss how to incorporate the
pair/molecule fluctuation directly into the final equations. We
also neglect here the possibility of a
non-zero-momentum pairing (the FFLO state \cite{4}, with the pair momentum $%
\mathbf{q\neq }0$). This is motivated by the fact that the FFLO
state is stable only within a narrow parameter window deep in the
BCS\ region \cite{8} and is absent in the recent $^{6}Li$
experiments \cite{2,3}.

With the mean-field approximation, one can then find out the
thermodynamical potential $\Omega =-T\ln[ $tr$\left(
e^{-H/T}\right)]$ at temperature $T$ (the Boltzmann constant is
taken to be $1$). It is given by the expression
\begin{eqnarray}
\Omega=-\Delta^2\mathcal{V}/U_T-T\sum_{\mathbf{k}}\{\ln[1+\exp(-\left|E_{\mathbf{k}\downarrow}\right|
/T)]\hspace{0.8cm}\nonumber\\+\ln[1+\exp(-E_{\mathbf{k}\uparrow}/T)]
-[\epsilon_{\mathbf{k}}
-\mu_{\downarrow}-\theta(E_{\mathbf{k}\downarrow})E_{\mathbf{k}\downarrow}]/T\},
\end{eqnarray}
where the parameter $U_{T}\equiv U-\alpha ^{2}/\left( \gamma -2\mu
_{\mathbf{ r}}\right) $, the quasi-particle excitation energy
$E_{\mathbf{k}\uparrow ,\downarrow }=\sqrt{(\epsilon
_{\mathbf{k}}-\mu _{\mathbf{r}})^{2}+\left| \Delta \right|
^{2}}\mp h,$ and the gap $\Delta =\alpha \langle
b_{0}\rangle/\sqrt{\mathcal{V}}
+U/\mathcal{V}\sum_{\mathbf{k}}\langle a_{-\mathbf{k},\downarrow
}a_{\mathbf{k},\uparrow
}\rangle =z\langle b_{0}\rangle/\sqrt{\mathcal{V}} $ ($z\equiv \alpha -U\left( \gamma -2\mu _{%
\mathbf{r}}\right) /\alpha $). The $\theta $-function is defined as $\theta
\left( x\right) =1$ for $x>0$ and $\theta \left( x\right) =0$ otherwise.
Note that different from the equal-population case, the excitation energies $%
E_{\mathbf{k\sigma }}$ are different for the $\mathbf{\sigma =}\uparrow
,\downarrow $ branches. Without loss of generality, we take $h<0$ so that $%
E_{\mathbf{k}\uparrow }>0$ always; while for a certain range of
$h$, the sign
of $E_{\mathbf{k}\downarrow }$ depends on the momentum $\mathbf{k}$ . When $%
E_{\mathbf{k}\downarrow }<0$, the atoms at that momentum are
actually not paired (and thus have no contribution to $\Delta $)
as pairing is energetically unfavorable (that is why we have
$\left| E_{\mathbf{k}\downarrow }\right| $ and $\theta
(E_{\mathbf{k}\downarrow })$ in Eq. (2)). This case corresponds to
the so-called breached pair state \cite{13}. From the variational
condition $\partial \Omega /\partial \left| \Delta \right| ^{2}=0
$, we get the gap equation
\begin{equation}
\frac{1}{U_{Tp}}=-\frac{1}{\mathcal{V}}\sum_{\mathbf{k}}(\frac{1-f(E_{\mathbf{k}\uparrow })-f(E_{%
\mathbf{k}\downarrow })}{E_{\mathbf{k}\uparrow }+E_{\mathbf{k}\downarrow }}-%
\frac{1}{2\epsilon _{\mathbf{k}}}),
\end{equation}
where the parameter $1/U_{Tp}\equiv 1/\left[ U_{p}-\alpha
_{p}^{2}/\left( \gamma _{p}-2\mu _{\mathbf{r}}\right) \right]
=1/U_{T}+(1/\mathcal{V})\sum_{\mathbf{k}}\left[ 1/\left( 2\epsilon
_{\mathbf{k}}\right) \right] $ (the latter equality comes from the
renormalization relation between $\gamma ,\alpha ,U$ and $\gamma
_{p},\alpha _{p},U_{p}$ \cite{11}). From the relation $\partial
\Omega /\partial \mu _{\sigma }=-n_{\mathbf{r}\sigma
}\mathcal{V}$, where $n_{\mathbf{r}\sigma }$ is the density of the
spin-$\sigma $ component, we get the number equations
\begin{equation}
n_{\mathbf{r}\sigma}
=\frac{1}{\mathcal{V}}\sum_{\mathbf{k}}[u_{\mathbf{k}
}^{2}f(E_{\mathbf{k},\sigma
})+v_{\mathbf{k}}^{2}f(-E_{\mathbf{k},-\sigma })]+ \frac{\left|
\Delta \right| ^{2}}{z^{2}},
\end{equation}
where the parameters
$u_{\mathbf{k}}^2=(E_{\mathbf{k}}+(\epsilon_{\mathbf{k}}-
\mu_{\mathbf{r}}))/2E_{\mathbf{k}}$,
$v_{\mathbf{k}}^2=(E_{\mathbf{k}}-(\epsilon_{\mathbf{k}}-
\mu_{\mathbf{r}}))/2E_{\mathbf{k}}$,
$E_{\mathbf{k}}=\sqrt{(\epsilon _{\mathbf{k}}-\mu
_{\mathbf{r}})^{2}+\left| \Delta \right| ^{2}}$, the fermi
distribution $f(E)\equiv 1/\left( 1+e^{E/T}\right) $, and we take
$-\uparrow=\downarrow$ and vice versa. The last term $ \left|
\Delta \right| ^{2}/z^{2}$ in $n_{\mathbf{r}\sigma }$ comes from
the contribution of the molecule (closed-channel) population,
which is small near a wide resonance. The atom densities
$n_{\mathbf{r}\uparrow }$ and $n_{ \mathbf{r}\downarrow }$ are
connected with the total atom number and the
population imbalance through $N=\int d^{3}\mathbf{r}\left( n_{\mathbf{r}%
\uparrow }+n_{\mathbf{r}\downarrow }\right) $, and $\beta =\int d^{3}\mathbf{%
r}\delta n_{\mathbf{r}}/N$ ($\delta n_{\mathbf{r}}\equiv n_{\mathbf{r}%
\downarrow }-n_{\mathbf{r}\uparrow }$).

The mean-field approach above neglects the pair/molecule
fluctuation. With this fluctuation taken into account, there will
be a non-condensed fraction of the pairs, which also contributes
to the gap for the fermionic quasiparticles \cite{11}. So the gap
now is replaced by $\left| \Delta \right| ^{2}=\left| \Delta
_{s}\right| ^{2}+\left| \Delta _{pg}\right| ^{2}$, where $\left|
\Delta _{s}\right| ^{2}=z^{2}\left| \langle b_{0}\rangle \right|
^{2}/\mathcal{V}$ and $\left| \Delta _{pg}\right|
^{2}=z^{2}m_{nc}/\mathcal{V}$ \cite{11}, with $\left| \langle
b_{0}\rangle \right| ^{2}$ and $m_{nc}$ representing the condensed
and the
non-condensed molecule numbers, respectively. With the interpretation of $%
\left| \Delta \right| $ as the total gap, the gap and the number
equations above
remain valid with the pair fluctuation (note that the contribution to $n_{%
\mathbf{r}\sigma }$ from the non-condensed molecules $m_{nc}$ is
automatically taken into account by the last term in Eqs. (4)).
If one
wants to break up $\Delta $ to find out the superfluid order parameter $%
\Delta _{s}$ and the pseudogap $\Delta _{pg}$, one needs to know
the dispersion relation for the pair/molecule excitations. Here,
to compare with the experiments \cite{2,3}, we only calculate the
total gap $\Delta $ and the atom density distributions
$n_{\mathbf{r}\sigma }$. For that purpose, the above gap and
number equations suffice, and we do not need to specify the pair
dispersion relation. Note that the total gap $\Delta $ is the
quantity directly measurable through the radio-frequency
spectroscopy \cite{15}.

The gap and the number equations above are in principle sufficient
to determine the distribution of $\left| \Delta \right| $ and
$n_{\mathbf{r}\sigma }$. However, different from the
equal-population case, the solution of this set of equations turns
out to be subtle, as they often give unstable phases or incorrect
results. To understand this, we examine the behavior of the
thermodynamical potential $\Omega $ under the variations of some
system parameters. Note that the gap equation should give an
optimal value of $\left| \Delta \right| $ which minimizes $\Omega
$. However, with a population imbalance, the potential $\Omega $
has a double well structure in many cases. In \ Fig. 1, we show
$\Omega $ as a function of $\left| \Delta \right| $ as we change
the chemical potential $\mu _{\mathbf{r}}$ (from the trap center
to the edge), the potential difference $h$ (with varying
population imbalance), the system temperature $T$, and the field
detuning $\gamma _{p}$. One can see that the double-well structure
shows up, in particular on the BCS\ side or in the
low-temperature, large-imbalance cases. Typically, one minimum of
the double-wells corresponds to $\left| \Delta \right| =0$ (the
normal state) and the other has a nonzero $\left| \Delta \right|
$. When $\Omega $ has a double-well structure, the gap equation
can give an incorrect result in several different ways: first, it
may pick up a solution corresponding to the maximum of $\Omega $
(the gap equation is satisfied at that point), which is obviously
not stable; second, it can choose the shallow well, which gives a
metastable but not the optimal configuration; and finally, the gap
equation (3) is usually not satisfied with $\left| \Delta \right|
=0$ even if the latter is the minimum of $\Omega $ (as $\partial
\Omega /\partial \left| \Delta \right| ^{2}\neq 0$ at that point).
In this case, the numerical program gives an incorrect result in
an uncontrollable way.

\begin{figure}[tbp]
\includegraphics{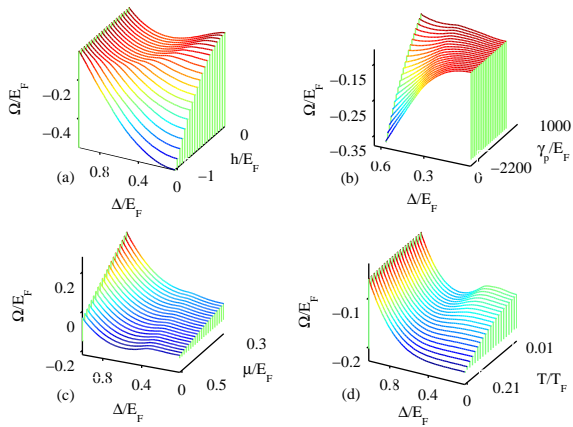}
\caption[Fig.1 ]{The thermodynamical potential $\Omega$ as a function of the
total gap $|\Delta| $ with varying (a) $h$ (the chemical potential
difference), (b) $\protect\gamma_p $ (the field detuning), (c) $\protect\mu $
(the chemical potential), and (d) $T$ (the temperature). We have used $%
T_{F}=E_{F}=k_{F}^{2}/2m$ as the unit of energy/tempeature, where $k_{F}=(3%
\protect\pi ^{2}n_{0})^{1/3}$ is a convenient inverse length scale
corresponding to a density $n_{0}=3\times 10^{13}cm^{-3}$, typical for the $%
{}^{6}Li$ experiments. The parameters $\protect\alpha_p, U_p$ take
the standard values for the ${}^{6}Li$ atoms with
$\protect\alpha_p \sqrt{n_0}=-0.6E_F$, and $U_p n_0 = -82 E_F $.
The other parameters for Figs.
(a-d) are given by (a)$\protect\mu =0.5E_{F}$, $\protect\gamma =0$, $%
T=0.01T_{F}$; (b) $\protect\mu=0.5E_{F}$, $h=-0.35E_{F}$, $T=0.01T_{F}$; (c)
$h=-0.35E_{F}$, $\protect\gamma =0$, $T=0.01T_{F}$; and (d) $\protect\mu %
=0.5E_{F}$, $h=-0.35E_{F}$, $\protect\gamma =0$.}
\end{figure}

To overcome the above problems, we always check the stability of
the solution by finding out all the minima of the thermodynamical
potential $\Omega $.
Note that although the metastable state at the bottom of a shallow well of $%
\Omega $ does not give the ground state configuration, it has a
finite relaxation time and may appear in real experiments under
certain circumstances (similar to the superheating or supercooling
phenomena in a classical phase transition). So, in that sense, the
state of the system may not be unique. To avoid this complexity,
in the following we assume the system to be in its equilibrium
configuration by choosing the solution corresponding to the global
minimum of the thermodynamical potential $\Omega $.

Through minimization of the thermodynamical potential, we have
calculated the gap and the density distributions for trapped
fermions with different magnetic field detunings, at both zero and
finite temperatures. Fig. 2 shows some typical results. To
summarize, we find that the gas is separated into several
different phases from the trap center to the edge, with the number
of phases and their boundary sensitive to the detuning $\gamma
_{p}$, the population imbalance $\beta $, and the temperature $T$.
At zero temperature and on the BEC side of the resonance with a
small population imbalance, there are typically three different
phases: the superfluid phase occupies the center of the trap,
where all the particles are paired with no population imbalance
(the SF phase). Further out from the center, when the pairing gap
becomes smaller than the chemical potential difference $\left|
h\right| $, there is a second phase, in which the particles are
paired except for a finite region of the momentum shell (the
breached pair phase \cite {6}, or in short, the BP\ phase). This
BP phase is characterized by a nonzero $\left| \Delta \right| $,
but with excessive fermions in the breached momentum shell, which
gives a finite population imbalance. In Fig. 2(a), the middle
region with $\left| \Delta \right| $ and $\delta n_{\mathbf{r}}$
both nonzero corresponds to such a phase. Towards the edge of the
trap, where the pairing gap has already vanished, only the normal
Fermi sea is left, with different Fermi surfaces for different
spin components. As the population imbalance increases, the SF
phase at the center of the trap depletes and yields to the
surrounding BP phase and the normal Fermi sea. At the resonance or
further out to the BCS side, this picture is different in that
there are now only two stable phases, the SF phase and the normal
phase, as shown in Fig. 2(b) (the BP\ phase in the middle loses
its stability). As the population imbalance increases and reaches
a critical value, the system undergoes a SF-normal phase
transition, and the trap is left with only the normal Fermi gas.

\begin{figure}[tbp]
\includegraphics{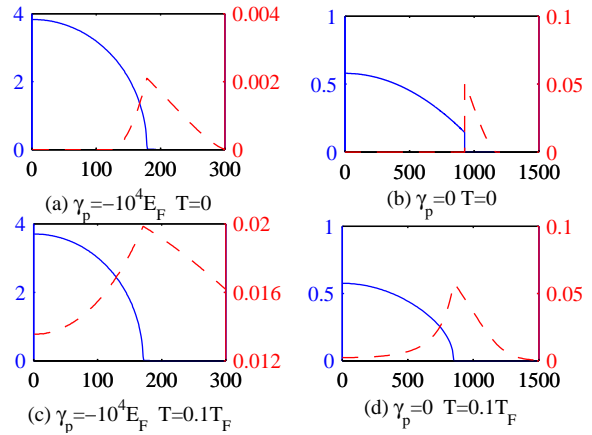}
\caption[Fig.2 ]{The gap ($|\Delta|$) and the differential density
($\left| \Delta \right| $ and $\delta n_{\mathbf{r}}$)
distributions for the trapped fermions, both at zero and at finite
temperature. The $x$-axis is the distance from the trap center, in
the unit of $k_Fr$. The solid curves represent $|\Delta|$ (the
left $y$-axis) and the dashed ones are for $\delta n_{\mathbf{r}}$
( the right $y$-axis). The other parameters are given by
(a)(c)$\protect\mu=-111.48E_F$, $h=-111.52E_F$; and (b)(d)
$\protect\mu=0.5E_F$, $h=-0.1E_F$. The corresponding population
imbalance ratio $\protect\beta$ and the total atom number $N$ are (a) $%
1\% $, $3.1\times10^5$; (b) $6\%$, $3.9\times 10^7$; (c) $69\%$,
$7.9\times 10^5$; and (d) $17\%$, $4.0\times 10^7$, respectively.}
\end{figure}

At finite temperature, the main difference is that there are
fermionic excitations in the SF phase, which carry population
imbalance of the spin components. As a result, it becomes easier
to satisfy the population imbalance constraint, as with a fixed
imbalance ratio $\beta $, the corresponding chemical potential
difference $\left| h\right| $ becomes significantly smaller. This
helps to stabilize the SF phase in the case of imbalanced
population. Because of this feature, we cannot use $\delta
n_{\mathbf{r}}$ to distinguish the SF and the BP phases any more,
so the boundary between them becomes obscure. The population
difference $\delta n_{\mathbf{r}}$ is peaked at the point where
the gap $\left| \Delta \right| $ vanishes, marking the only
distinguishable phase separation between the paired phase (SF or
BP) and the normal phase (see Fig. 3c and 3d). At higher
temperatures, the paired phase in the trap shrinks and finally
disappears at the pair disassociation temperature $T^{\ast }$ (the
superfluidity should disappear before that with $T_{c}<T^{\ast }$
\cite{11}). In Fig. 2(c) and (d), we show the distribution of
$\left| \Delta \right| $ and $\delta n_{\mathbf{r}}$ at a typical
temperature of $0.1E_{F}$, for the BEC and the resonance-BCS
sides, respectively.

\begin{figure}[tbp]
\includegraphics{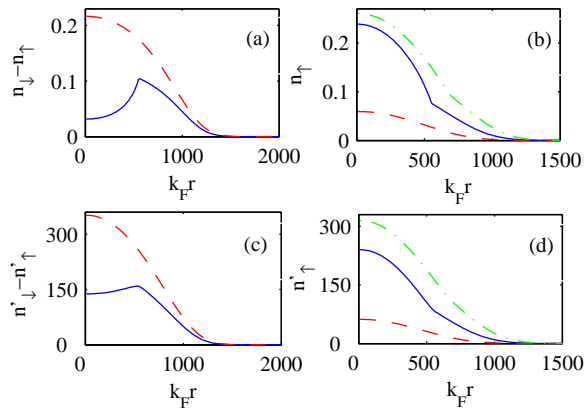}
\caption[Fig.3 ]{The atom density distributions in a trap,
calculated with
parameters corresponding to the recent MIT experiment \cite{2} with $%
T\sim 0.1E_F$, and $\protect\gamma\sim 2100E_F$
($B-B_0\sim56.3G$). The solid,
the dashed, and the dash-dot curves correspond to the imbalance ratio $%
\protect\beta \sim46\%$ (the corresponding $\protect\mu=0.5E_F$, $h=-0.12E_F$%
), $\protect\beta \sim86\%$ (the corresponding $\protect\mu=0.43E_F $, $%
h=-0.23E_F$), and $\protect\beta =0$ ($\protect\mu=0.5E_F $, $%
h=0$) respectively, with the total atom number $N\sim
2.7\times10^7$ in all the cases. Figs. (c,d) show the column
integrated density distributions corresponding to Figs. (a,b).}
\end{figure}

Finally, to compare with the recent MIT experiment \cite{2}, we
calculate the density distributions $n_{\mathbf{r}\sigma }$, with
all the parameters roughly the same as those in the experiment
(the experimental temperature may have some uncertainties). The
results are plotted in Fig. 3. It shows a pretty good agreement
with the experimental data (Fig. 3 of Ref. \cite{2}), at least
semi-quantitatively. In particular, the calculation shows that the
distribution $\delta n_{\mathbf{r}}$ has a dip at the trap center
with a
moderate population imbalance ($\beta =46\%$); while the dip disappears when $%
\beta $ becomes large ($86\%$), which agrees exactly with the experimental
findings.

In summary, we have studied the effects of trap and temperature on
fermions across a wide Feshbach resonance with population
imbalance. We propose to directly minimize the thermodynamical
potential in order to overcome the instability problem associated
with the solution of the gap equation. We establish a general
phase separation picture for the trapped fermions across the whole
resonance region, at both zero and finite temperatures. We have
also compared our calculation with a recent experiment, and
recovered some of the main experimental findings.

\textbf{Note added}: Upon completion of this work, two preprints
[P. Pieri, G.C. Strinati, cond-mat/0512354; J. Kinnunen, L.M.
Jensen, P. Torma, cond-mat/0512556] appeared on arxiv, where a
similar problem are investigated at zero temperature with
different theoretical approaches.

We thank Jason Ho, Martin Zwierlein, Wolfgang Ketterle, and Kathy Levin for
helpful discussions. This work was supported by the NSF award (0431476), the
ARDA under ARO contracts, and the A. P. Sloan Foundation.


\begin{thebibliography}{99}
\bibitem{13}  This phase has various names in the literature, see Refs. \cite
{6,5,8}. We follow the terminology of Ref. \cite{6}.
\end{thebibliography}
\end{document}